# Vacuum Calorimetry in Exploding Wire Studies

**Mario Rabinowitz**

*Department of Physics*
*Washington State University*
*Pullman, Washington*

Inquiries to: *Armor Research*
*715 Lakemead Way, Redwood City, CA 94062*
*Mario715@earthlink.net*

**Abstract**

A method of using vacuum calorimetry as a means of determining directly the energy deposited in an electrically exploded wire is presented. This energy determination is compared with that given by the time integral of the product of voltage and current. A definite reproducible pattern of the explosion products is deposited on the walls of the calorimeter, which may be used as a means of understanding the behavior of the wire during the explosion.

**Introduction**

Thin metal wires are exploded by means of the rapid deliverance of electrical energy from a capacitor discharged in a time interval of the order of microseconds. In previous studies of exploding wires, the experimenters have measured by electrical methods the energy given to the wire [l, 2, 3]. This has either been done by means of a simultaneous measurement of current, I, and voltage, V, or by use of the measured value of the current together with a knowledge of the circuit parameters of capacitance, C, self-inductance, L, and resistance, R, of the circuit exclusive of the wire. In the former case, the energy given to the wire is given by

$$\int_0^t IV\,dt.$$

In the latter case by

$$\tfrac{1}{2}CV_0^2 - \left[\tfrac{1}{2}\frac{(Q_0 - \int_0^t I\,dt)^2}{C} + \tfrac{1}{2}LI^2 + R\int_0^t I^2\,dt\right].$$

(Vo) and $Q_0$ are respectively the initial voltage and charge on the capacitor.)

To study the possibilities of a new technique for making a measurement of the energy given to the wire, vacuum calorimeters were designed and used in making a direct measurement of the energy given to the wire. The calorimeters were in the form of thin aluminum cylinders enclosing the coaxial wire that was to be exploded. The pressure in the calorimeter, and in its external environment inside a vacuum chamber, was of the order of 10-5 torr.

There are three main reasons for operating the calorimeter under vacuum conditions. First, if the calorimeter is to be strong enough to withstand the strong shock generated by a wire explosion in air, the mass of the calorimeter would have to be so great that the resulting temperature change would be too small to be significant. Secondly, if the calorimeter system were not isolated, the conduction and convection losses for the small temperature changes obtained (~10 oC would not be tolerable. The third reason is that when the metals are exploded in air at atmospheric pressure, they tend to form metallic oxides in a chemical reaction that is exothermic. Unless the percentage of oxide formed were known, one would have no way of correcting for the energy thus released.

Therefore, the simplest way to overcome these difficulties is by exploding the wire in an evacuated calorimeter which is itself in a vacuum chamber (Fig. 1). Because the pressure pulse on the calorimeter is much smaller, one may use the minimum amount of material for structural strength. Essentially the only energy loss is due to radiation from the calorimeter, and this is quite small. By operating at sufficiently low pressures (10-5 torr), there are not enough oxygen molecules present to allow oxidation to be a problem. When the copper explosion residue was submitted for X-ray crystallographic analysis, copper oxide was undetectable.

Simultaneous current and voltage measurements were made on the exploding wire with each calorimetric determination so that the energy measured by the calorimeter could be compared with the electrical measurement of the energy given by $\int_0^t IV dt.$ The current, I, and the voltage, V, were measured as functions of time by standard methods [1, 2].

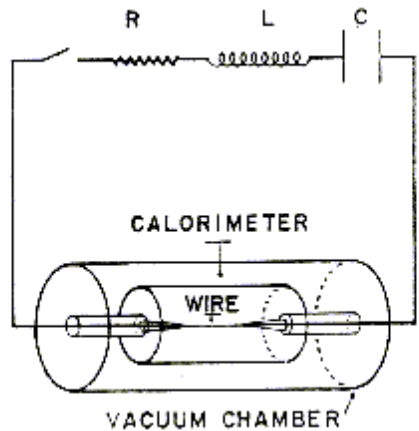

Fig. 1. Schematic diagram of the circuit with the calorimeter inside the vacuum chamber. The capacitance ranged from I to 15 µF, charged to a maximum voltage of 50,000 V.

## Design and calibration of the calorimeters

The calorimeters were designed to have a minimal energy equivalent of mass, m times specific heat, c, so that the temperature change, $\Delta T$. was maximized for a given energy input Q. The calorimeter equation is $\Delta T = Q/mc$. In order to keep the mass small. the calorimeter which is in the form of a right circular cylinder, was made of minimal dimensions of diameter, length, and thickness sufficient to withstand the pressure pulse from the explosion, and to prevent electrical breakdown from the electrodes to the walls.

The calorimeters were made of aluminum, with energy equivalents of about 115 J/oC, where the differences are due to dimensional variations. A typical calorimeter is shown in Fig. 2. The wall thickness is approximately 0.02 in, the diameter is 3.5 in., and the length is 4 in. The calorimeter is supported at each end by a teflon insulated electrode which protrudes into the calorimeter. A small hole in the base of the calorimeter near the electrode hole facilitated evacuation. A correction was made for the hole and the presence of the electrodes inside the calorimeter.

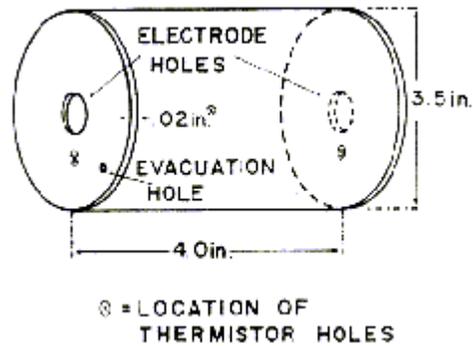

Location of Thermistor Holes

Fig. 2. A typical calorimiter.

Three thermistors were placed in shallow holes in the walls of each calorimeter to measure the temperature change. They all indicated the same temperature rise to better than +/- 0.1 C. The thermistors were small Fenwal bead thermistors BC3261 (diameter 0.007 in, and diameter of leads 0.001 in) arid GA45J I (diameter 0.04 in and leads 0.004 in). These thermistors were chosen because of their desirable properties of extremely small mass, fast response time, and precision reproducibility. The resistance of the thermistors as a function of temperature was determined to better than 0.1 C accuracy by means of a controlled temperature oven and precision wheatstone bridge 2.

The calorimeters were calibrated by electrically heating a resistor that replaced the exploding wire in the evacuated calorimeter. The amount of energy put in, was thus known to within I per cent accuracy. The essential difference between this heating and the heat supplied by a wire explosion is that the resistor remained in contact with the electrodes during the whole time of this experiment requiring a correction for end loss by heat conduction along the electrical leads : whereas. when a wire is exploded. it is in contact with the electrodes for only about a microsecond. The electrodes are insulated from the calorimeter. so in this case there is essentially no heat loss to the leads. The energy equivalent values thus obtained agreed to better than 10 per cent with those computed directly from the product of mass and specific heat after correction was made for the presence of the electrodes and evacuation holes.

**Comparison of calorimetric measurements and electrical energy**

The energy measured by the calorimeters was consistently less than the electrical energy given by Ee = $\int_0^t IV dt.$ . The energy measured by the calorimeter, Ec, had an experimental error less than +/-10 per cent, and the accuracy of Ee, was within +/-15 per cent as given by the uncertainties in the current and voltage measurement. However, within the error limits, Ec + *10* per cent was always in fair agreement with Ee - 15 per cent. Typical values are : Ec = *700* J +/- 10 per cent and *Ee - 890 J = +/-* 15 per cent for an energy storage in the capacitor of 3000 J, using a 2 in. long No. *28* copper wire.

Though the two energy determinations are in fair agreement within the error limits, it may actually be correct for the calorimeter to measure a smaller energy than that measured electrically. Of the energy given to the wire not all goes into kinetic energy of the wire particles ; a significant amount of energy is used in vaporizing the metal. It is quite possible that all or part of the heat of vaporization is not given back to the calorimeter. The explosion products may remain in a higher energy state than that of the wire. Therefore, the calorimeter would essentially measure only the kinetic energy given to the wire arid part of' the energy used in vaporization.

For the purpose of checking into the difference in energy determinations. 2 in. long No. 28 copper wires of mass 0.036 g were used in a series of tests where the energy input to the wire was varied. Two methods of varying the energy given to the calorimeter were used. One was that of charging the capacitor to a smaller initial voltage, i.e. reducing the total energy storage. The other was that of using a shunting switch to divert the current from the exploding wire after a given amount of energy had been delivered to it (Fig. 3).

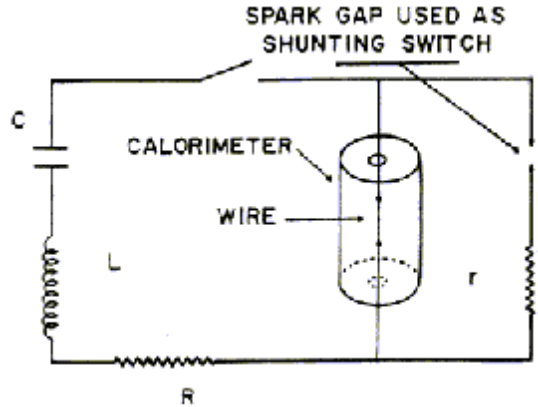

Fig. 3. Shunting switch added to original circuit.

The shunting switch is an air spark gap. When the electric field across the switch is high enough to cause the gap to break down, its resistance becomes very small compared with that of the wire, and the current is shunted away from the wire stopping the energy input to it. Using the shunting switch, energies as small as 40J were given to the calorimeter. On such occasions, the wire did not vaporize, but broke up into tiny pieces. The pieces were successively smaller as more energy was supplied to the wire.

When the wire was vaporized, the energy difference, $E_e - E_c$, was consistently of the order of the heat of vaporization of the wire which is approximately 140 J. Similar results were obtained with wires of other materials such as aluminum, and steel, but the evidence is not conclusive as yet.

**Copper deposit pattern**
During the series of shots using the 2 in. No. 28 copper wire, a definite pattern of the explosion products was deposited on the inside of the calorimeter after each shot. A similar pattern was formed even when the wire was not completely vaporized. In this case, the copper was deposited on the walls in a pattern consisting of long thin filaments. The appearance suggests that molten bits of the wire had hit the walls at very small glancing angles producing streaks in a direction roughly perpendicular to the axis of the wire. This would tend to indicate that the explosion particles do not move in a straight line path from the wire-but rather that their trajectory is curved; possibly indirect evidence of ionization. Until now, it has been commonly accepted that the explosion particles move in a radial direction from the wire in a vacuum environment.

The deposit of the completely vaporized wire was even more striking. Not only was the over-all pattern present, but there were striations in the pattern forming a fine structure. This is clearly shown in Fig. 4, a photograph of a copper wire explosion pattern. This particular deposit was obtained by putting mylar film next to the cylinder wall of the calorimeter to catch the deposit. (The long edges formed the circumference of the cylinder.)

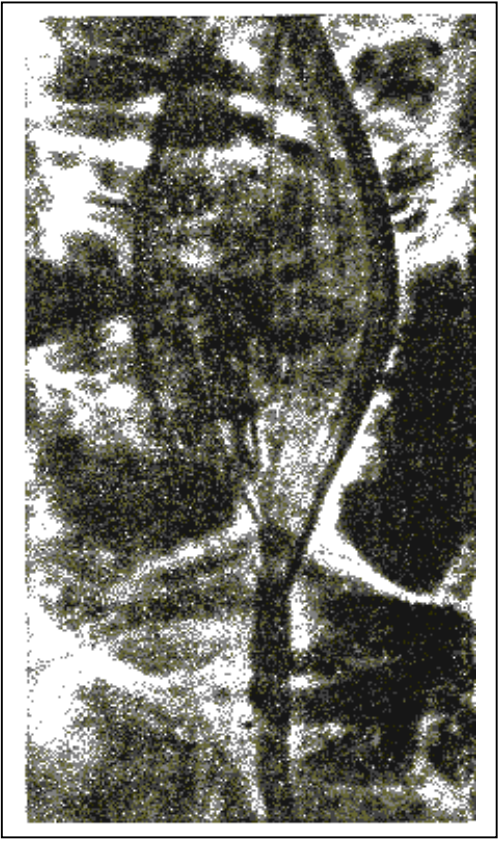

Fig. 4. Photograph of the copper wire explosion pattern. The length of the wire is approximately the same as the distance across the larger maximum. The long edges formed the circumference of the cylinder.

Since one might expect to observe a more or less uniform deposit because the calorimeter cylinder is coaxial with the wire, it was surprising to observe the over-all pattern with the fine structure. The pattern does not occur for explosions of wires in a coaxial cylinder at atmospheric pressure and even at a pressure of only 10-5 torr when the diameter of the enclosing cylinder was increased to 6 in, the deposit did not form a pattern. This might indicate that the pattern results from a strong focusing action of the electromagnetic field of the discharge current upon the exploding wire.

One possible explanation, intended only to be heuristic, is that the field causes the initially straight wire to bow in the form of an arc before it vaporizes. Then as vaporization proceeds, the wire receives an acceleration in the direction of bowing. This would account for a greater condensate concentration on one side of the cylindrical wall than the other. The leading edge of the explosion would encounter relatively few collisions. The following portion would suffer more collisions due to slower moving particles of the explosion. It would also be less influenced by the electromagnetic field, which would by then be in the process of diminishing. This scattering of the trailing particles might then account for the amorphous background superimposed upon the pattern. The striations may be indicative of a pinch instability in the exploding wire.

A more rigorous analysis would have to explain the over-all pattern shape together with the fine structure striations. This might prove to be a difficult but rewarding pursuit.

## Conclusion

Though vacuum calorimetry affords one of the most direct methods of determining the energy deposited in an electrically exploded wire, the difference between the energy values as determined electrically and by calorimetry must be better understood before it can become a truly reliable and accurate instrument.

The explosion deposit pattern represents a potential tool in understanding the behavior of the wire during the explosion process. It appears that a theoretical analysis of the pattern is indeed a challenging problem.

## Acknowledgments

The author wishes to thank Dr. D. V. Keller and J. R. Thomas for their helpful suggestions. The experimental work was done at the Boeing Company, Seattle, Washington. Thanks are also due to Professor E. E. Donaldson for his encouragement.